\documentclass[showpacs,preprintnumbers,amsmath,amssymb]{revtex4}
 \usepackage{natbib}
\usepackage{graphicx}
\usepackage{dcolumn}
\usepackage{bm}
\begin{document}
\title{Effect of exchange interaction on superparamagnetic relaxation}
\author{Hamid Kachkachi}
\email{kachkach@physique.uvsq.fr}
\affiliation{
  Lab. de Magn\'{e}tisme et d'Optique, Univ. de Versailles St. Quentin,
45 av. des Etats-Unis, 78035 Versailles CEDEX, France\\
}
\begin{abstract}
We use Langer's approach to calculate the reaction rate of a system
of two (classical) spins interacting via the exchange coupling $J$ in a
magnetic field $H$, with uniaxial anisotropy of constant $K$.
We find a particular value of the exchange coupling, that is $j\equiv J/K =
j_c\equiv 1-h^2$, where $h\equiv H/2K$, which separates two regimes
corresponding to a two-stage and one-stage switching.
For $j\gg j_c$ the N\'eel-Brown result for the one-spin problem is
recovered.
\end{abstract}
\maketitle
\section{Introduction}
Due to their high coercivity, single-domain nanoparticles exhibit
long-range stability of the magnetization and thereby that of the
information stored in recording media. The
storage density may be increased by using very small
particles, but then surface effects become dominant and affect the
magnetization relaxation.
Therefore, one of our aims is to include surface effects in the
calculation of the relaxation time of the particle
magnetization. However, this requires a microscopic approach to account for
the local environment inside the particle, and thus include microscopic
interactions such as spin-spin exchange, in addition to anisotropy and
applied field.
Unfortunately, this leads to a rather difficult task owing to the large
number of degrees of freedom which hinders any attempt to analyze the
energyscape.
For this reason, inter alia, calculations of the reversal time of the
magnetization of fine single-domain ferromagnetic particles, initiated by
N\'{e}el \cite{Neel}, and set firmly in the context of the theory of
stochastic processes by Brown \cite{Brown}, have invariably
ignored all kind of interactions and included only the internal
anisotropy of the particle, the random field due to thermal fluctuations, and
the Zeeman term.

In 1968 Langer \cite{Langer} came up with a general approach to the
calculation of the reaction rate of a system with $N$ degrees of
freedom, which is a powerful method of attack on the problem of
interaction.
Within this approach the problem of calculating the relaxation time
for a multi-dimensional process is reduced to solving a steady-state
Fokker-Planck equation (SSFPE) in the immediate neighborhood of the saddle point
that the system crosses as it goes from a metastable state to another state of
greater stability.
The idea is that a steady-state situation can be set up by continuously
replenishing the metastable state at a rate equal to the rate at which it is
leaking across the activation energy barrier.
A brief review of the main steps and application to the
problem at hand are given after Eq.~(\ref{Gamma_tsp}), see also \cite{Braun} for
uniform and non-uniform magnetization, and \cite{CoffeyGaraninMcCarthy} for
comparison with Kramers' theory.
However, Langer's approach is only valid in the limit of
intermediate-to-high damping (IHD) because of the inherent assumption that the
potential energy in the vicinity of the saddle point may be approximated by its
second-order Taylor expansion.
The result for small damping fails because the region of deviation from the
Maxwell-Boltzmann distribution set up in the well extends far
beyond the narrow region at the top of the barrier.

The problem associated with the generalization of Brown's theory to
include interactions is rather difficult and can in general be solved only
numerically.
But before attacking this problem, one needs to understand the
effect of exchange interaction on the relaxation time of the minimal
system, i.e., for a pair of atomic spins coupled via exchange interaction,
including of course the usual magneto-crystalline anisotropy and Zeeman terms.
Besides, in the case of exchange interaction, this is the unique non-trivial
step towards the above-mentioned generalization, where analytical expressions
can be obtained for the relaxation time.
It is the purpose of this work to solve this problem within Langer's
approach and to compare with the N\'eel-Brown result for the one-spin
problem.
As a byproduct, we will show that Langer's quadratic approximation at the
saddle point fails when the exchange coupling assumes a ``critical value"
even in the IHD limit.
In this case, finiteness of the relaxation rate requires a higher-order
expansion of the energy near the saddle point.
The generalization to a multi-spin particle will be briefly discussed at the
end.

We finally mention that the effect of dipolar interactions on the relaxation
time has been considered by a few authors.
In Refs.~\cite{assembly} approximate analytical expressions were obtained for
the relaxation time in various situations of assemblies of magnetic moments.
In Ref.~\cite{ChantrellLyberatos} a pair of coupled dipoles was dealt with using
the (numerical) Langevin approach.

\section{Statement of the problem and analysis of the energyscape}
We consider a system of two exchange-coupled spins with the Hamiltonian
\begin{eqnarray}
&&{\cal H} =-\frac{j}{2}\vec{S}_{1}\cdot\vec{S}_{2}-\frac{1}{2}\left[
(\vec{S}_{1}\cdot\vec{e}_{1})^{2}+(\vec{S}_{2}\cdot\vec{e}_{2})^{2}\right]
-\vec{h}\cdot(\vec{S}_{1}+\vec{S}_{2}), \nonumber \\
&&j\equiv J/K,\, h\equiv H/2K  \label{HTS}
\end{eqnarray}
where $J>0$ is the exchange coupling, $K>0$ the
anisotropy constant, $H$ the applied magnetic field, and $h$
the reduced field, i.e., $0\leq h < 1$. $\vec{e}_i$ are uniaxial
anisotropy unit vectors. Here we restrict ourselves to the case
$\vec{h}\parallel\vec{e}_i, i=1,2$.
%
%
Owing to the
symmetry of this system with respect to rotations around the easy axis, the
number of variables reduces to three, $\theta_1, \theta_2,
\varphi\equiv\varphi_1-\varphi_2$.

\begin{figure}[floatfix]
\includegraphics[angle=-90, width=13cm]{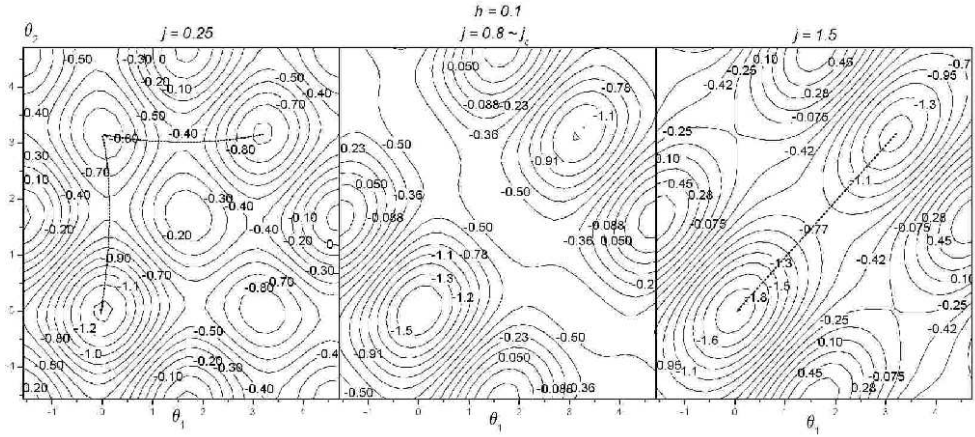}
\caption{\label{enscape}
Energyscape from Eq.~(\ref{HTS}) ($\varphi=0$) for $j=0.25, 0.8, 1.5$ and
$h=0.1$. The arrows indicate the switching paths.}
\end{figure}

Now we apply Langer's approach to the energy (\ref{HTS}) and study the
relaxation rate as a function of the exchange coupling $j$.
We first analyze the energyscape in Fig.~\ref{enscape}.
%
%
The absolute minimum of the energy (\ref{HTS}) corresponds to the
ferromagnetic order along the easy axis,
\begin{equation}
(0,0,\varphi);\quad e_{ss}^{(0)}=-\frac{j}{2}-2h-1,  \label{sfm}
\end{equation}
where henceforth $e^{(0)}$ denotes the energy of the state.
One metastable state corresponds to a ferromagnetic order opposite to
the field
\begin{equation}
(\pi,\pi;\varphi);\quad e_{m_1}^{(0)}=-
\frac{j}{2}+2h-1.  \label{mfm}
\end{equation}
There is also the metastable state of anti-ferromagnetic order,
\begin{equation}\label{afm}
(0,\pi;\varphi)\,\mbox{or}\,(\pi,0;\varphi);\quad
e_{m_2}^{(0)}=\frac{j}{2}-1.
\end{equation}
As to saddle points, we find that their number and loci crucially depend on
the exchange coupling constant $j$. More precisely, for $j
>j_c\equiv 1-h^2$, there is a single saddle point given by
\begin{equation}
(\cos\theta_1=\cos\theta_2 = -h; \varphi);\quad
e_{s}^{(0)}=-\frac{j}{2}+h^{2},  \label{SP0}
\end{equation}
whereas for $j < j_c$ there are two saddle points given by
\begin{eqnarray}\label{SP21}
\cos\theta_{1,2}^{\varepsilon}= \frac{1}{2}\left( -h-a^{\varepsilon}\pm
\sqrt{\Delta^{\varepsilon}}\right)\equiv X^{\varepsilon}_{\pm}; \,\, \varphi, &  &
\end{eqnarray}
where $\varepsilon=\pm$ and
$
a^{\varepsilon} = \varepsilon\sqrt{1-j}, \, b^{\varepsilon} =
\varepsilon\sqrt{1+(j/2)^2/(1-j)}, \,
\Delta^{\varepsilon}=(h+a^{\varepsilon})^2+2j-4b^{\varepsilon} h,
$
%
%
with energy,
\begin{equation}\label{enj<jc}
e_{\varepsilon}^{(0)}=-\frac{j}{2}\sqrt{(1+b^{\varepsilon}
h-\frac{j}{2})^2-(a^{\varepsilon}+ h)^2} - \frac{j}{2}(1+b^{\varepsilon}
h-\frac{j}{2})+ \frac{1}{2}(h^2-(a^{\varepsilon})^2+2b^{\varepsilon} h).
\end{equation}
At $j=j_c$ the saddle point (\ref{SP21}) with
$\varepsilon=+$ merges with the saddle point (\ref{SP0}), while that with
$\varepsilon=-$ merges with the metastable state (\ref{afm}),
see Fig.~\ref{enscape} central panel.

Starting with both spins aligned in the metastable
state (\ref{mfm}) with $\theta_{1,2}=\pi$, if $j < j_c$ one of the
two spins crosses the saddle point (\ref{SP21})($\varepsilon=+$) into the state
(\ref{afm}) by reversing its direction.
Then the second spin follows through the second saddle point
(\ref{SP21})($\varepsilon=-$), of lower energy due to the
exchange coupling (see Fig.~\ref{rate} (left)), ending up in the stable
state (\ref{sfm}).
In Fig.~\ref{enscape} (left) the path is indicated by a pair of curved arrows.
There are actually two such paths corresponding to the two-fold symmetry of
the problem owing to the full identity of the two spins.
Note that when the first spin starts to switch and arrives at
$\theta_1\sim\pi/2$, the second spin has $\theta_2\lesssim\pi$ (hence the
curved arrows in Fig.~\ref{enscape}), which suggests that in the switching
process of the first spin, the position of the second spin
undergoes some fluctuations creating a small transverse field, and when
$\theta_1=0$ the second spin goes back to the position $\theta_2=\pi$
before it proceeds to switch in turn.
The successive switching of the two spins through the corresponding saddle
points is a sequential two-step process
\footnote{\label{footnoteChLyb}Similarly, it was found in
Ref.~\cite{ChantrellLyberatos} that the reversal of the two dipoles considered
is a two-stage process with an intermediary metastable antiparallel state.}, so
the relaxation rates for $j < j_c$ add up inverse-wise.
In the case $j > j_c$ the two spins cross the
unique saddle point (\ref{SP0}) to go from the metastable state (\ref{mfm}) into
the stable one (\ref{sfm}) in a single step, see Fig.~\ref{enscape} (right)
where the path is indicated by a single straight arrow.
There is the symmetry $\pm\theta^{(s)}$,
which leads to a factor of 2 in the relaxation rate.
Therefore, if we denote by $\Gamma_{j\leq j_c}^+$, $\Gamma_{j\leq j_c}^-$,
and $\Gamma_{j \geq j_c}$ the respective relaxation rates, the relaxation rate
of the two-spin system is given by
\begin{equation}\label{Gamma_tsp}
\Gamma = \left\{
\begin{array}{ll}
2\Gamma_{j\leq j_c}^+ \Gamma_{j\leq j_c}^-/(\Gamma_{j\leq
j_c}^+ + \Gamma_{j\leq j_c}^-),  & \mbox{if $j \leq j_c$} \\
2\Gamma_{j \geq j_c}, & \mbox{if $j \geq j_c$}.
\end{array}
\right.
\end{equation}
%
%

In Langer's approach the $\Gamma$s are obtained from
the SSFPE linearized around each saddle point, using the fluctuating variables
$\eta_i=(t_i,p_i)$ where
$t_i\equiv\theta_i-\theta_i^s,\,p_i\equiv\varphi_i-\varphi^s_i , \, i=1,2$, or
more adequately the ``canonical" variables
$\psi_n=(\xi_{\pm},\zeta_{\pm})$ with $\xi_{\pm}=(t_1\pm
t_2)/\sqrt{2},\,\zeta_{\pm}=(p_1\pm p_2)/\sqrt{2}$,
in which the energy Hessian is diagonal.
The deterministic dynamics of the system is governed by the Landau-Lifshitz
equations, which upon linearization near the saddle point, read
\begin{equation}\label{LinLLE}
\left\{
\begin{array}{ll}
\partial_t\,t_i=-\partial_{p_i}{\cal H}_2-\alpha \partial_{t_i}{\cal H}_2\\
\partial_t\,p_i=-\alpha\partial_{p_i}{\cal H}_2 + \partial_{t_i}{\cal H}_2,\quad
i=1,2
\end{array}\right.
\end{equation}
where ${\cal H}_{2}$ is the quadratic approximation of the energy (\ref{HTS})
near the saddle point and $\alpha$ is the damping parameter.
In matrix form Eqs.~(\ref{LinLLE}) become
$
\partial_{t}\eta_{i}=\sum_{j}M_{ij}\,\partial_{\eta_{j}}{\cal H}_{2}
$,
where $M$ is the dynamic matrix containing the precessional and
dissipative parts.
Rather than investigating the stochastic trajectories
$\eta_{i}(t)$ that arise by adding a noise term in Eqs.~(\ref{LinLLE}), Langer
concentrates on the distribution function $\rho(\{\eta\},t)$ as the probability
that the system is found in the configuration $\{\eta\}$ at time $t$. The time
evolution of $\rho$ is governed by the Fokker-Planck equation (FPE)
$
\partial_t\rho +\sum_i\partial_{\eta_i}J_{i}=0
$,
which is a continuity equation with the probability current $J_{i}$ given, near
the saddle point, by
$
J_{i}=-\sum _{j}M_{ij}\left(\partial_{\eta_j}{\cal H}_2 +
(1/k_BT)\partial_{\eta_j}\right)\rho.
$
In order to calculate the nucleation rate, one must solve the
FPE. A particular solution is obtained at equilibrium and is
given by the Maxwell-Boltzmann distribution $\rho_{eq} = \exp (-\beta {\cal
H}\left\{\eta\right\})/Z_0$, which corresponds to zero current, $J_{i}=0$.
However, what is really needed is a finite probability current flowing across
the saddle point.
In fact, instead of solving the time-dependent FPE,
Langer solves the SSFPE $\partial_t\rho = 0$ near the saddle point.
The steady-state situation may be realized by imposing the boundary conditions:
$\rho\simeq\rho_{eq}$ near the metastable state and $\rho\simeq 0$ beyond the
saddle point.
Then, the problem of calculating the escape rate reduces to the calculation
of the total current by integrating the probability current $J_i$ over a
surface through the saddle point.
This led Langer to his famous expression for
the relaxation rate which is valid in the IHD limit.

However, since the escape rate is simply given by the ratio of the total current
through the saddle point to the number of particles in the metastable state,
Langer's result for the escape rate can in fact be achieved by only computing
the energy-Hessian eigenvalues near the saddle points and metastable states,
from which one then infers the partition function $\tilde{Z_s}$ of the system
restricted to the region around the saddle point where the energy-Hessian
negative eigenvalue is (formally)\footnote{See Ref.~\cite{Langer} for a rigorous
derivation} taken with absolute value, and the partition function $Z_m$ of the
region around the metastable state.
When computing these partition functions, one has to identify and take care of
each Goldstone mode, that is a massless mode or zero-energy fluctuation
associated with a continuous unbroken global symmetry.
Finally, one computes the unique \footnote{Indeed, if the saddle
point is to describe the nucleating fluctuation, there must be exactly one
direction of motion away from the saddle point in which the solution of the
equations of motion of the modes $\psi_n$ is unstable \cite{Langer}.}
negative eigenvalue $\kappa$ of the SSFPE corresponding to the unstable mode at
the saddle point as the negative eigenvalue of the dynamic matrix
$\tilde{M}_{mn}=-\lambda_n(DMD^T)_{mn}$, where the $\lambda_n$'s are the
eigenvalues of the Hessian at the saddle point and $D$ is the transformation
matrix from $\eta_i$ to $\psi_n$.

Consequently, Langer's final expression for the escape rate is rewritten
in the following somewhat more practical form
\begin{equation}\label{Gammapr}
\Gamma=\frac{\left|\kappa\right|}{2\pi}\frac{\tilde{Z_s}}{Z_m},
\end{equation}
where $\left|\kappa\right|$ is the attempt frequency which contains the
damping parameter $\alpha$.

\section{Relaxation rate of the two-spin system}
Now, we give the different relaxation rates for $j > j_c, j < j_c$, and
$j\simeq j_c$.

{\bf Relaxation rate for $j > j_c$}: The quadratic expansion of the energy
(\ref{HTS}) at the saddle point (\ref{SP0}) reads,
\begin{equation}\label{QEj>j_c}
H^{(2)}_s=e_s^{(0)}-
\frac{j_c}{2}\xi_{+}^2+\frac{j-j_c}{2}\xi_{-}^2+\frac{jj_c}{2}
\zeta_{-}^2,
\end{equation}
with zero eigenvalue for the $\zeta_+$ mode, that is the Goldstone
mode associated with the rotation around the easy axis. $e_s^{(0)}$ is
given in Eq.~(\ref{SP0}).
The negative eigenvalue of the SSFPE, corresponding to the unstable mode,
is $\kappa =-\alpha j_c$.

The partition function at the saddle point $Z_{s}=\int d\Omega_1
d\Omega_2 e^{-\beta H_{s}},$
where $d\Omega_i=\sin\theta_i d\theta_i d\varphi_i,\, i=1,2$ is calculated
by changing to the variables $\xi_{\pm},\zeta_{\pm}$, setting
$\sin\theta_i\simeq \sqrt{j_c}$ (see Eq. (\ref{SP0})) in the
integration measure, and finally computing the Gaussian integrals. Hence,
\begin{equation}
\label{Z_sp}
\tilde{Z}_{s}=2\pi \left(
\frac{2\pi}{\beta}\right)^{3/2}\frac{1}{j\sqrt{1-j_c/j}}e^{-\beta
e^{(0)}_{s}},
\end{equation}
where we have formally replaced the negative eigenvalue $(-j_c)$ by
its absolute value.
The partition function $Z_m$ at the metastable state
(\ref{mfm}) is computed by expanding the energy up to $2^{\mbox{nd}}$
order, leading to
\begin{equation}\label{Z_m}
Z_{m}=e^{-\beta e^{(0)}_{m1}}\left( \frac{2\pi }{\beta }\right)
^{2}\frac{1}{(1-h)(j+1-h)}.
\end{equation}

Using Eq.~(\ref{Gammapr}) and inserting the symmetry factor of $2$, the
relaxation rate for $j>j_c$ reads
\begin{equation}\label{Gamma_tsp_j>jc}
\Gamma_{j>j_c}=2\alpha \sqrt{\frac{\beta }{2\pi
}}(1-h^{2})(1-h)\frac{1+(1-h)/j}{\sqrt{1-j_c/j}}\times e^{-\beta
\Delta e^{(0)}}, \quad \Delta e^{(0)}=(1-h)^{2}.
\end{equation}

For $j\rightarrow \infty$ (\ref{Gamma_tsp_j>jc}) tends to the N\'eel-Brown
result,
\begin{equation}\label{NB}
\Gamma_{j > j_c}\rightarrow 2\alpha
\sqrt{\frac{\beta }{2\pi }}(1-h^2)(1-h)\,\,e^{-\beta(1-h)^{2}},
\end{equation}
for the relaxation rate of one rigid pair of spins with a barrier height
twice that of one spin.
Note, however, that this convergence is very slow so that
$\Gamma_{j > j_c}$ remains above the N\'eel-Brown result and only
merges with the latter for $j\gtrsim 10$.

{\bf Relaxation rate for $j < j_c$}: Here we are faced with a semi-analytical
case since the attempt frequencies $|\kappa^{\varepsilon}|$ (for
$\varepsilon=\pm$) cannot be obtained in a closed form and are thus computed
numerically.
On the other hand, following the same procedure as for $j > j_c$ we obtain
the relaxation rate for $\varepsilon=\pm$,
%
\begin{equation}\label{gamma_jleqjc}
\Gamma_{j < j_c}^{\varepsilon}=\sqrt{\frac{\beta}{\pi}}|\kappa^{\varepsilon}|
\sqrt{\frac{P^{\varepsilon}}{j}}\frac{N^{\varepsilon}}{\sqrt{R^{\varepsilon}_+
R^{\varepsilon}_- + jQ^{\varepsilon}(R^{\varepsilon}_+
+ R^{\varepsilon}_-)}} \, e^{-\beta\Delta e^{(0)}_{\varepsilon}},
\end{equation}
%
where $\Delta e^{(0)}_{\varepsilon}$ is the barrier height given by
the energy (\ref{enj<jc}) measured with respect to (\ref{mfm}) or
(\ref{afm}) for $\varepsilon=+,-$ respectively, and (see Eq.~(\ref{SP21})
et seq. for notation)
%
$
P^{\varepsilon}=\sqrt{(1+b^{\varepsilon}h-j/2)^2-(a^{\varepsilon}+h)^2},
Q^{\varepsilon}=b^{\varepsilon}h-j/2+P^{\varepsilon},
R^{\varepsilon}_{\pm} = -1+X^{\varepsilon}_{\pm}(2X^{\varepsilon}_{\pm} +
h), N^+ = (1-h)(j+1-h), N^-=j_c-j.
$
%
The limit of the relaxation rate (\ref{gamma_jleqjc})
when $j\longrightarrow 0$ is just the N\'eel-Brown result for one spin.
Indeed, the product of the last two factors in the prefactor tend to
$(1-h)/\sqrt{2}$, the attempt frequencies tend to $\alpha j_c=\alpha
(1-h^2)$, and the energy barriers $\Delta
e^{(0)}_\varepsilon\rightarrow (1-h)^2/2$.

Note that in the present regime of $j<j_c$ the large value of anisotropy
has not changed the temperature dependence of the individual relaxation
rates, i.e., $\Gamma_{j < j_c}^{\varepsilon}$, with $\varepsilon = \pm$.
This is due to the fact that $1/\sqrt{T}$ appears in the prefactor each
time there is a continuously degenerate class of saddle points
\cite{Braun}, which is indeed the case for $j>j_c$ and $j<j_c$ with
$\varepsilon = +$ and $\varepsilon = -$.
However, anisotropy do affect the temperature dependence of the relaxation
rate of the two-spin system, since for $j < j_c$ there are two
saddle points bringing each a factor $1/\sqrt{T}$, see the first line in
Eq.~(\ref{Gamma_tsp}).

\begin{figure}[floatfix]
\includegraphics[angle=-90, width=14cm]{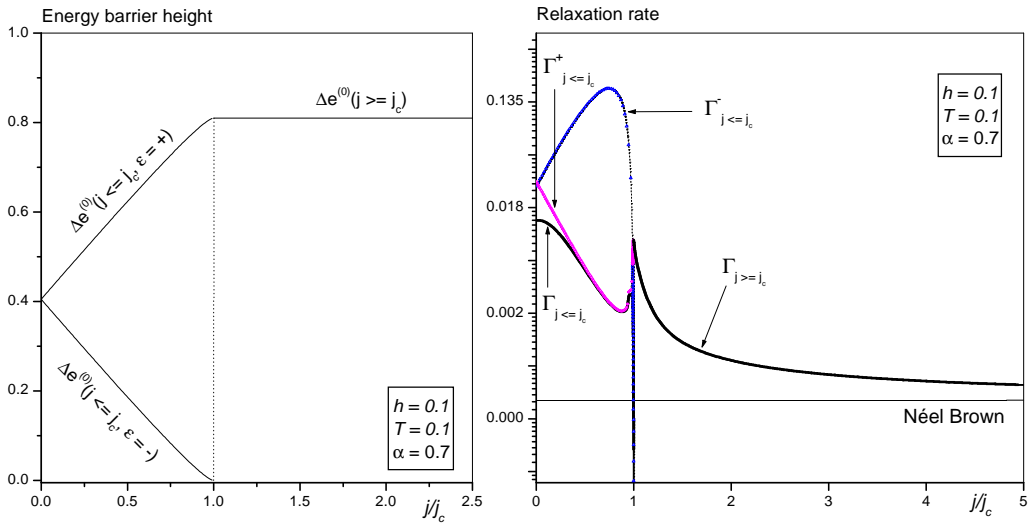}
\vspace{-2cm}
\caption{\label{rate}
Left: Energy barrier as a function of $j$.
Right: Relaxation rate (in logarithmic scale) of the two-spin system in the case
of intermediate-to-high damping. The straight horizontal line is the
N\'eel-Brown result with the barrier height taken twice that of a single spin.}
\end{figure}

%
{\bf The case of $j\simeq j_c$}: When $j$ approaches $j_c$ either from above or
from below, more Hessian eigenvalues (in addition to $\lambda_{\zeta^+}$)
vanish, rendering the saddle point rather flat and thus leading to a divergent relaxation
rate.
Indeed, for $j\simeq j_c$ the
relaxation rate (\ref{Gamma_tsp_j>jc}) diverges, which clearly shows that
Langer's approach which uses a quadratic approximation for the energy at
the saddle point, e.g., Eq.~(\ref{QEj>j_c}), fails in this case.
The remedy is to push the energy expansion to the $6^{th}-$order in the variable
$\xi_-$ (since $\lambda_{\xi_-}=j-j_c\rightarrow 0$ as $j\rightarrow
j_c$), i.e., 
\begin{equation}\label{E6}
\delta e_s = e_s-e_s^{(0)}\simeq
\frac{\lambda_{\xi_-}}{2}\xi_-^2+\frac{c}{4}\xi_-^4+\frac{d}{6}\xi_-^6,
\end{equation}
where 
$c=(1-j-7h^2/4)/3<0, \,d=(j-1+31h^2/16)/30>0$.
%
%
Then, the contribution $\sqrt{2\pi/\beta \lambda_{\xi_-}}$ of the mode $\xi_-$ to
the relaxation rate (\ref{Gamma_tsp_j>jc}) must be replaced
by$\int_{-\infty}^{\infty}d\xi_- e^{-\beta\delta e}$, upon which the
divergence of $\Gamma_{j>j_c}$ is cut off (see Fig.~\ref{rate} (right)).
%
%
Similarly, when $j$ approaches $j_c$ from below,
for $\varepsilon=-$ the eigenvalue $\lambda_{\xi^-}$ vanishes at
$j=j_c$, upon which the saddle point (\ref{SP21}) ($\varepsilon=-$) merges
with the state (\ref{afm}) and thereby the partition functions
$\tilde{Z}_s$ and $Z_m$ tend to infinity. However, as $Z_m\propto
1/(j_c-j)$ increases much faster than $\tilde{Z}_s$, and
$|\kappa^-|\rightarrow 0$, $\Gamma_{j < j_c}^-$ tends to zero as
$j\rightarrow j_c$.
On the other hand, for $\varepsilon=+$, both $\lambda_{\xi^-}$ and
$\lambda_{\zeta^-}$ vanish leading to a divergent relaxation rate, since
now $\tilde{Z}_s$ diverges but $Z_m$ in Eq.~(\ref{Z_m})
remains finite for the metastable state (\ref{mfm}) is well defined.
Indeed, as $j\rightarrow j_c$, the relaxation rate
$\Gamma_{j < j_c}^+$ goes over to the result in Eq.~(\ref{Gamma_tsp_j>jc})
upon making the change $j\leftrightarrow j_c$, and taking account of the
symmetry factor.  
In this case, the divergence at the point $j=j_c$ cannot be cut off
by expanding the energy beyond the $2^{\mbox{nd}}$ order and $\Gamma_{j <
j_c}^+$ is simply cut off at the point where it joins $\Gamma_{j > j_c}$
taking account of Eq.~(\ref{E6}).
\section{Discussion and conclusion}
In Fig.~\ref{rate} (right) we plot ($ln$ of) the relaxation rate of the two-spin
system as defined in Eq. (\ref{Gamma_tsp}), under the condition of
IHD, in which Langer's approach is valid, that the reduced barrier height
$\beta\Delta e^{(0)}\gg 1$ and $\alpha\beta\Delta
e^{(0)}>1$ \cite{DG}. We also plot separately both relaxation rates for
$j\leq j_c$.
We see that the relaxation rate of the two-spin system contains two
unconnected branches corresponding to the two regimes, $j < j_c$ and
$j > j_c$, the bridging of which would require a more sophisticated
approach.
Fig.~\ref{rate} (right) also shows that as $j$ increases, but $j \ll j_c$, the
relaxation rate $\Gamma_{j\leq j_c}^+$ decreases because the switching of the
first spin is hindered by the (ferromagnetic) exchange coupling. While
$\Gamma_{j\leq j_c}^-$ is an increasing function of $j$ with a faster rate,
since now the exchange coupling works in favor of the switching of the second
spin. %
This is also illustrated by the evolution of the energy barrier height in
Fig.~\ref{rate} (left).
As $j$ approaches $j_c$ from below, the relaxation rate $\Gamma_{j\leq
j_c}^+$ tends to $\Gamma_{j \geq j_c}$ because the respective saddle points
merge at $j=j_c$.
Whereas $\Gamma_{j\leq j_c}^-$ goes to zero since the corresponding saddle
point merges with the antiferromagnetic state that is no longer
accessible to the system.
For $j\geq j_c$, as $j$ increases the
minimum (\ref{sfm}), the metastable state (\ref{mfm}) and the saddle point
(\ref{SP0}) merge together, which means that the system is
found in an ``energy groove'' along the direction $\theta_1=\theta_2$
because the eigenvalue $\lambda_{\xi_-}$ corresponding to the mode
$\theta_1-\theta_2$ becomes very large, and thus the escape rate decreases
and eventually reaches the N\'eel-Brown value at large $j$.

The present study has helped us understand the effect of exchange coupling
on the relaxation rate of the two-spin system, and will be very
useful for the generalization to multiple-spin small particles at least for
small deviations from collinearity, where it has been shown \cite{DGHK}
that the surface contribution to the macroscopic energy has a simple cubic
anisotropy.
However, this generalization can only be performed using numerical
techniques. This is now attempted by the help of the ridge method
\cite{IonovaCarter} for probing the energyscape and locating the
saddle points, and by the (Onsager-Machlup) path integrals \cite{Berkov} for
determining the most probable paths connecting a metastable state to a more
stable state.
\begin{acknowledgments}
The author is greatly indebted to D.A. Garanin for instructive suggestions.
He also thanks W.T. Coffey and D.J. McCarthy for helpful discussions at an early
stage of this work.
\end{acknowledgments}

\end{document}